\documentclass[onecolumn,12pt]{article}%
\usepackage[latin1]{inputenc}
\usepackage{graphicx}
\usepackage{amsmath}
\usepackage{cite}%
\usepackage{amsfonts}%
\usepackage{amssymb}
\newcommand \be{\begin{equation}}
\newcommand \ee{\end{equation}}

\begin{document}

\title{Modeling the Electrostatic Potential of Disks with Arbitrary Radial Charge Profiles}
\author{J. Ricardo de Sousa$^{a}$, Marcio Gomes$^{b}$, and Orion Ciftja$^{c}$\\$^{a}$Departamento de F\'{\i}sica e National Institute of Science and \\Technology for Complex Systems, \\Universidade Federal do Amazonas, \\3000, Japiim, 69077-000, Manaus-AM, Brazil.\\$^{b}$Instituto Federal de Educa\c{c}\~{a}o, Ci\^{e}ncia e Tecnologia do Amazonas,\\Av. 7 de setembro, 1975 - Centro, 69020-120, Manaus-AM, Brazil. \\$^{c}$Department of Physics, Prairie View A\&M University, Prairie View\\Texas 77446, USA}
\date{\today    }
\maketitle

\begin{abstract}
In this work, we present an analytical study of the electrostatic potential
generated by a charged disk with a surface charge distribution 
that possesses radial axial symmetry. 
We express the potential in cylindrical coordinates and 
apply a Bessel function representation for the
Coulomb kernel to simplify its calculation. We consider a broad family of
surface charge densities and derive an exact expression for the potential
based on a classical integral identity involving Bessel functions. This
general formulation includes several important special cases such as the
uniformly charged disk and edge-concentrated charge density distributions. We
explore how variations in various parameters affect the resulting potential,
thereby illustrating the influence of radial charge modulation. 
It is found that the potential undergoes a qualitative change of behavior
when the chosen set of parameters is varied. 
The results are analytically elegant and computationally efficient as all functions
involved are implemented in standard scientific computing libraries.

\end{abstract}

\section{Introduction}

The analysis of charge distributions with some form of symmetry is a
cornerstone of electrostatics, offering both conceptual clarity and practical
advantages in solving physical problems. Symmetry principles, in fact,
permeate virtually every domain of physics, where they serve not only as
guiding tools for formulating theories but also as powerful means of
simplifying complex mathematical descriptions. In many cases, the presence of
symmetry allows for a dramatic reduction in the computational effort required
to solve field equations, transforming what would otherwise be intractable
problems into analytically or numerically solvable ones. In electrostatics,
configurations that exhibit cylindrical or spherical symmetry enable the
adoption of coordinate systems that align naturally with the physical system.
This alignment, in turn, facilitates the use of mathematical techniques such
as separation of variables, integral transforms, and expansions in orthogonal
functions---including Legendre polynomials and Bessel functions. These
techniques are often impractical or even impossible to apply in the absence of
symmetry. As a result, identifying and exploiting symmetrical properties not
only provides deeper insight into the underlying physics but also leads to
more elegant and efficient solutions.




For the cases of cylindrical symmetry, such as uniformly charged rings and
disks, the analysis becomes significantly intricate when one looks to
determine the electrostatic potential at off-axis points. The analysis
generally leads to complicated integral representations involving special
functions such as \textit{complete elliptic integrals}, \textit{Bessel
functions} and \textit{Legendre polynomials}. Unlike the spherical case, where
symmetry can be fully exploited through Gauss's law or straightforward
integration, the lack of spherical symmetry in cylindrical systems prevents
such simplifications. As a result, exact analytical solutions are usually
confined to highly symmetric points, particularly along the axis of symmetry
as extensively covered in standard physics textbooks\cite{6,7}. Despite the
apparent simplifications associated with the cylindrical symmetry, the full
mathematical treatment of the electrostatic potential at an arbitrary point in
three-dimensional (3D) space is far from trivial~\cite{8,9,10,11,12,13}. Such
a treatment typically requires a deeper engagement with special functions that
are less familiar to a general audience. This mathematical complexity adds
both theoretical richness and pedagogical value to the problem, offering a
more nuanced understanding of the interplay between geometry, symmetry and
electrostatic behavior.

Among the available techniques used, one of the most direct and conceptually
transparent methods for computing the electrostatic potential produced by a
uniformly charged ring or disk---particularly at points not lying along the
axis of symmetry---is to perform a direct integration starting from the
fundamental electrostatic potential formula and then use known mathematical
transformations from the literature. This method, grounded in Coulomb's law,
leads naturally to integral expressions involving special functions,
especially the complete elliptic integrals of the first kind. Although these
integrals are more advanced than those typically encountered, their use in
this context is both justified and well established. Detailed derivations and
results can be found throughout the scientific literature, attesting to the
robustness and versatility of this
approach\cite{el1,el2,el3,el4,el5,el6,el7,el8}. The problem of determining the
electrostatic potential generated by a uniformly charged disk is of
significant importance, not only from a mathematical standpoint within
electrostatics, but also due to its wide range of applications in physical
systems that exhibit disk-like geometries.
For example, the modeling and analysis of gravitational fields generated by
disk-like mass distributions have received considerable attention in the
scientific literature\cite{g1,g2,g3,g4,g5,g6,g7,g8,g9,g10}. This interest
stems from the fundamental influence of these mass configurations on orbital
dynamics, the formation of planetary systems, and the overall structure of
galaxies.
Another practical use of the charged disk model lies in its effectiveness as
an approximation for the electrostatic confinement of electrons in a variety
of nanostructured materials and devices, such as nanowires and carbon
nanotubes. In many of these systems, the spatial distribution of electric
charge or induced dipoles gives rise to electrostatic potentials that exhibit
axial symmetry and radial profiles closely resembling those produced by a
uniformly charged disk\cite{n1,n2}. As a result, the disk model provides a
valuable theoretical framework for analyzing electron behavior in nanoscale
systems, where confinement effects and electrostatic interactions play a
critical role in determining electronic, optical, and transport properties. By
capturing the essential features of the potential landscape, this model
supports the development of analytical and computational techniques for
predicting the performance and behavior of nanodevices under realistic
physical conditions.

Furthermore, the charged disk model has proven valuable in fields such as
colloid physical chemistry where it is used to describe the structure and
behavior of colloidal systems as well as biological
environments\cite{b1,b2,b3,b4}. Specifically, many biological
macromolecules---including proteins, DNA strands and micelles---as well as
electrolyte solutions containing large, charged particles, can, under certain
conditions, be effectively approximated by charge distributions exhibiting
disk-like geometries. This approximation facilitates the understanding of
electrostatic interactions and stability within these complex systems,
providing insights into processes such as molecular assembly, colloidal
stability, and biological function. Another significant area of application
lies within condensed matter physics where the charged disk model has been
utilized to describe phenomena related to the integer/fractional quantum Hall
effect. In this context, the model aids in representing the positive
neutralizing baxkground in two-dimensional (2D) electronic systems, playing a
crucial role in advancing the understanding of electrostatic interactions in
these environments\cite{f1,f2,f3,f4,f5}. By providing a simplified yet
effective geometric framework, the charged disk approximation helps elucidate
how charge distributions and screening effects influence electron behavior,
contributing to deeper insights into quantum electronic phases and the
emergent collective properties of low-dimensional materials.
%

In this work, we will study the electrostatic propertied of a charged disk
with surface charge density that possesses radial symmetry. The mathemarical
analysis is based on Bessel functions and their properties. Our approach will
yield exact expressions for a wide range of surface charge densities with
axial symmetry, thus providing a versatile and rigorous framework for
understanding these electrostatic configurations. The structure of this paper
is as follows: Section 2 introduces the physical system under study and
outlines the theoretical framework employed to calculate the electrostatic
potential. In Section 3, we analyze and interpret the key findings that emerge
from our methodology. Lastly, Section 4 provides a summary of the conclusions
drawn from this work and highlights potential avenues for further investigation.

\section{Model and Methodology}

The model investigated in this work consists of a thin, circular disk with
radius, $a$ carrying a non-uniform surface charge distribution with axial
symmetry. The net charge in the disk is denoted by, $Q$.
The disk lies in the $xy$-plane and is centered at the origin of the
coordinate system. Our primary goal is to determine the resulting
electrostatic potential at an arbitrary point in space, both on and off the
disk plane. This configuration serves as a fundamental model for exploring the
effects of radially varying charge distributions on the spatial behavior of
electrostatic fields and potentials. Such scenarios are commonly encountered
in a wide range of physical systems including thin conductive films, charged
biological membranes, and planar electrodes with tailored surface charge profiles.

To facilitate the analysis, we employ cylindrical coordinates where the 3D
position vector is expressed as $\overrightarrow{r}=x\overrightarrow
{i}+y\overrightarrow{j}+z\overrightarrow{k}=\overrightarrow{\rho
}+z\overrightarrow{k}=(\overrightarrow{\rho},z)$ where $\overrightarrow{\rho
}=x\overrightarrow{i}+y\overrightarrow{j}$ is a 2D vector lying in the plane
of the disk and $\overrightarrow{i}$, $\overrightarrow{j}$ and
$\overrightarrow{k}$ are the unit vectors in the directions of the Cartesian
axes $x$, $y$, and $z$, respectively. In cylindrical coordinates, the
Cartesian components of a point in the plane are given by $x=\rho\,
\cos\varphi$ and $y=\rho\, \sin\varphi$ where $\rho=\left|  \overrightarrow
{\rho}\right|  \geq0$ is the radial distance from the axis of symmetry (the
$z-$axis) and $\varphi\in\left[  0,2\pi\right]  $ is the azimuthal angle.

To compute the electrostatic potential, $\Phi(\vec{r})$ generated by a disk
carrying a non-uniform axial symmetric surface charge distribution,
$\sigma(\rho)$, we begin with the general expression for the potential due to
a continuous surface charge distribution, as found in standard texts on
electrostatics\cite{6,7,8,9,10,11,12,13}:
\begin{equation}
\Phi(\overrightarrow{r})=\frac{1}{4\pi\varepsilon_{o}}\int_{\text{Disk}
}\frac{dq}{\left|  \overrightarrow{r}-\overrightarrow{r}^{\prime}\right|  }
\ , \label{q1}%
\end{equation}
where $\overrightarrow{r}^{\prime}=(\overrightarrow{\rho}^{\prime},0)$
represents a position vector in the plane of the disk, $\overrightarrow
{r}=(\overrightarrow{\rho},z)$ is the arbitrary point in space, $dq=\sigma
(\rho^{\prime})\rho^{\prime}d\rho^{\prime}d\varphi^{\prime}$ is the
infinitesimal charge element at $\overrightarrow{r}^{\prime}$ and $"Disk"$
represents the disk region of integration.

The kernel $\left|  \overrightarrow{r}-\overrightarrow{r}^{\prime}\right|
^{-1}$ in the integrand can be expanded using the cylindrical addition theorem
for Bessel functions\cite{18}, which is given by the following identity:%
\begin{equation}
\frac{1}{\left|  \overrightarrow{r}-\overrightarrow{r}^{\prime}\right|  }%
=\sum\limits_{m=-\infty}^{\infty}\int\limits_{0}^{\infty}e^{i\,m\,(\varphi
-\varphi^{\prime})}J_{m}(k\,\rho)J_{m}(k\,\rho^{\prime})e^{-k\left|
z-z^{\prime}\right|  }\,dk\ , \label{q2}%
\end{equation}
where $J_{m}(x)$ denotes the Bessel function of the first kind of order $m$
and $i=\sqrt{-1}$ is the imaginary unit. Due to the axial symmetry of the
problem---specifically, because the surface charge density $\sigma(\rho)$
depends only on the radial coordinate $\rho$ --- the resulting electrostatic
potential is also independent of the azimuthal angle $\varphi$, and can
therefore be expressed as $\Phi(\vec{r})=\Phi(\rho,z)$. Substituting the
expansion from Eq.(\ref{q2}) into the integral expression in Eq.(\ref{q1}), we
obtain the following representation for the electrostatic potential in
cylindrical coordinates:
\begin{equation}
\Phi(\rho,z)=\frac{1}{4\pi\varepsilon_{o}}\sum\limits_{m=-\infty}^{\infty
}e^{i\,m\,\varphi}\int\limits_{0}^{\infty}J_{m}(k\,\rho)e^{-k\left|  z\right|
}dk\int\limits_{0}^{2\pi}e^{-i\,m\,\varphi^{\prime}}d\varphi^{\prime}%
\int\limits_{0}^{a}J_{m}(k\,\rho^{\prime})\sigma(\rho^{\prime})\rho^{\prime
}d\rho^{\prime}\ . \label{neweq3}%
\end{equation}
The angular integral over $\varphi^{\prime}$ yields $2\pi\delta_{m,0}$, due to
the orthogonality of the complex exponentials over the interval $\left[
0,2\pi\right]  $. This result implies that only the $m=0$ term contributes to
the sum, significantly simplifying the expression. As a result, we arrive at
the general expression for the potential due to an arbitrary radial surface
charge distribution:%
\begin{equation}
\Phi(\rho,z)=\frac{1}{2\varepsilon_{o}}\int\limits_{0}^{\infty}J_{0}%
(k\rho)\left[  \int\limits_{0}^{a}J_{0}(k\rho^{\prime})\sigma(\rho^{\prime
})\rho^{\prime}d\rho^{\prime}\right]  e^{-k\left|  z\right|  }dk\ , \label{q3}%
\end{equation}
which holds for any axi-symmetric surface charge density, $\sigma(\rho
^{\prime})$.

At this juncture, we remark that it is possible to derive a general analytical
expression for the electrostatic potential in Eq. (\ref{q3}), for the case of
a circular disk carrying a non-uniform surface charge distribution whose
radial profile is described by the functional form:
\begin{equation}
\sigma(u)=\sigma_{o}\left(  1-u^{2}\right)  ^{\nu}u^{p} \ , \label{q4}%
\end{equation}
where $u=\rho/a$ is a 2D dimensionless radial coordinate expressed in terms of
the disk radius, $a$ while $\nu$ and $p$ are real parameters that characterize
the shape of the charge distribution.

Specifically speaking, the exponent, $\nu$ controls the behavior of the
surface charge density near the disk's edge ($u\rightarrow1$) potentially
introducing an edge singularity. On the other hand, the exponent $p$ controls
the behavior of the charge density near the center of the disk ($u\rightarrow
0$) possibly leading to a central singularity or depletion depending on its
value. The prefactor $\sigma_{o}$ is a normalization constant which can be
chosen in such a way as to ensure that the total charge on the disk equals the
prescribed value of $Q$.
This generalized class of surface charge distributions allows for a flexible
modeling framework that encompasses a wide range of physically relevant
scenarios, including uniform, edge-concentrated and center-enhanced charge
profiles. This class of charge distributions is particularly advantageous due
to its versatility in modeling a broad spectrum of physically relevant
scenarios. By appropriately selecting the values of the parameters $\nu$ and
$p$, one can reproduce several distinct charge configurations. For example,
the case $\nu=0$ and $p=0$ corresponds to a uniform surface charge
distribution. Values of $\nu<0$ yield edge-enhanced profiles, characterized by
a higher concentration of charge near the disk's boundary while $p>0$ produces
center-enhanced distributions, in which the charge density is more
concentrated near the center of the disk. As such, the functional family
defined by Eq.(\ref{q4}) provides a convenient and analytically tractable
framework for investigating how spatial variations in surface charge density
influence the structure of the resulting electrostatic potential. This
flexibility makes this choice particularly suitable for applications involving
non-uniform charge profiles such as those arising in the modeling of charged
membranes, electrodes with engineered surface properties, or electrostatic
problems involving axial symmetry.

To determine the normalization constant $\sigma_{o}$, we impose the condition
that the total charge contained over the surface of the disk must be fixed to
the value of $Q$. The total charge is obtained by integrating the surface
charge density $\sigma(\rho^{\prime})$ over the entire area of the disk. In
cylindrical coordinates, this is expressed as:
\begin{equation}
\int\limits_{0}^{a}\sigma(\rho^{\prime})2\pi\rho^{\prime}d\rho^{\prime}=2\pi
a^{2}\sigma_{o}\int\limits_{0}^{1}\left(  1-u^{\prime2}\right)  ^{\nu
}u^{\prime p+1}du^{\prime}=Q\ , \label{total}%
\end{equation}
where we have changed variables to the dimensionless radial coordinate,
$u^{\prime}=\rho^{\prime}/a$. This integral yields a finite result for
$\nu>-1$ and $p>-1$, ensuring the physical admissibility of the charge
distribution. The resulting integral can be expressed in terms of the
\textit{Euler beta function}, defined as\cite{18}:
\begin{equation}
B(m+1,n+1)=\int\limits_{0}^{1}t^{m}\left(  1-t\right)  ^{n}dt=\frac{m!n!}%
{(m+n+1)!}\ , \label{q5}%
\end{equation}
so that the result becomes:
\begin{equation}
\sigma_{o}\frac{\pi a^{2}(p/2)!\nu!}{(\nu+p/2+1)!}=Q\ . \label{q6}%
\end{equation}
Solving this equation for $\sigma_{o}$, we obtain:
\begin{equation}
\sigma_{o}=\frac{Q}{\pi a^{2}}\frac{(\nu+p/2+1)!}{(p/2)!\nu!}\ . \label{q7}%
\end{equation}
This expression ensures that the total charge of the disk matches the
prescribed value, $Q$ for any given pair of parameters $\nu>-1$ and $p>-1$,
thereby allowing the model to accommodate a wide variety of radial charge
distributions. The beta function, $B(p/2+1,\nu+1)$, which appears frequently
in mathematical physics, serves as a normalization factor and captures the
integral properties of the non-uniform distribution. As such, it provides a
compact and elegant means of accommodating a wide class of charge profiles
within a unified analytical framework.

By substituting Eqs. (\ref{q4}) and (\ref{q7}) into the general expression for
the electrostatic potential, Eq. (\ref{q3}), we arrive at an integral 
representation of the potential:
%
\begin{equation}
\Phi_{p,\nu}(\rho,z)=\frac{(\nu+p/2+1)!V_{o}}{(p/2)!\nu!}
\int\limits_{0}^{1} \left(  1-u^{\prime \, 2}\right)  ^{\nu}
\left[  \int\limits_{0}^{\infty}
J_{0}(q \, u)J_{0}(q \, u^{\prime})e^{-q \, \left|  \overline{z}\right|  } \, dq \right]
{u^{\prime}}^{\, p+1} \, du^{\prime}, \label{q8}%
\end{equation}
where 
$q=k \, a$ is a dimensionless dummy variable,
$\overline{z}=z/a$ is a dimensionless vertical coordinate
and  $V_{o}=Q/(2 \, \pi \, \varepsilon_{o} \, a)$. 
Note that, from now on, we denote the potential as $\Phi_{p,\nu}(\rho,z)$ 
in order to draw attention to the two parameters $p$ and $\nu$.
The expression in Eq.(\ref{q8}) describes the electrostatic
potential generated by an arbitrary radial surface charge distribution
with the functional form given from Eq.(\ref{q4}).

Although the integral presented in Eq.(\ref{q8}) may initially appear
complicated due to the presence of Bessel functions in the integrand, it can
be treated analytically by applying well-established identities from the
theory of special functions. In particular, we can use a classical identity
involving the product of two Bessel functions of the same order multiplied by
a decaying exponential factor. This integral is listed as formula
\textbf{6.612.3} in the renowned reference by Gradshteyn and Ryzhik\cite{be2},
and it states:
\begin{equation}
\int\limits_{0}^{\infty}J_{\mu}(xu)J_{\mu}(xu^{\prime})e^{-x\left|
\overline{z}\right|  }dx=\frac{1}{\pi\sqrt{u^{\prime}u}}Q_{\mu-\frac{1}{2}%
}\left(  Z\right) \ , 
\label{q8a}
\end{equation}
where 
$Q_{\mu}(Z)$ is known as the \textit{Legendre function of the second kind}
and we have introduced the dimensionless auxiliary variable:
%
\begin{equation}
Z=\frac{u^{\prime2}+u^{2}+\overline{z}^{2}}{2uu^{\prime}} \ .
\label{parZ}
\end{equation}
This variable satisfies $Z\geq1$ for all $u\geq0$, 
which ensures that the Legendre function, $Q_{\mu}(Z)$ is well-defined.
%
For specific values of the parameter $\mu$, such as integers or half-integers, 
$Q_{\mu}(Z)$ admits explicit representations in terms of logarithmic expressions 
or known definite integrals. One of the most widely used forms is its integral representation:

\begin{equation}
Q_{\mu}(Z)=\int\limits_{0}^{\infty}\frac{d\varphi}{\left[  Z+\sqrt{Z^{2}%
-1}\cosh(\varphi)\right]  ^{\mu+1}} \ . 
\label{q8b}
\end{equation}
This representation is explicitly listed as formula \textbf{8.822.2} in
Ref.\cite{be2} and is widely recognized in the mathematical literature as a
standard expression for the Legendre function of the second kind.
The integral form above is particularly advantageous both for theoretical analysis
and for computational purposes. 
It enables a direct numerical evaluation of $Q_{\mu}(Z)$ 
using standard quadrature methods since the integrand contains
only elementary functions such as the square root and hyperbolic cosine. 
We use the identity in Eq. (\ref{q8a}) for the $\mu=0$ case 
in order to calculate the integral expression in Eq. (\ref{q8}).
As a result, we obtain a simplified analytical expression 
for the potential generated at an arbitrary point is space:
\begin{equation}
\Phi_{p,\nu}(\rho,z)=\frac{(\nu+p/2+1)!V_{o}}{\pi\sqrt{u}(p/2)!\nu!}%
\int\limits_{0}^{1}\left(  1-u^{\prime2}\right)  ^{\nu}Q_{-\frac{1}{2}}\left(
Z\right)  u^{\prime \, p+1/2}du^{\prime} \ . 
\label{q8c}
\end{equation}
By adapting formula \textbf{560.01} from Ref.\cite{be3} to our case, we
find that the Legendre function of the second kind, $Q_{-\frac{1}{2}}\left(
Z\right)  $, can be expressed directly in terms of the complete elliptic
integral of the first kind, $K(m)$, namely,
\begin{equation}
Q_{-\frac{1}{2}}\left(  Z\right)  =\sqrt{m}\, K(m) \ , 
\label{q8d}
\end{equation}
where the modulus parameter, $m$ is defined as:
\begin{equation}
m=\frac{2}{1+Z}=\frac{4u^{\prime}u}{\left(  u+u^{\prime}\right)
^{2}+\overline{z}^{2}} \ . 
\label{q8e}%
\end{equation}
Finally, by inserting the result from Eq. (\ref{q8d}) into the general
expression in Eq. (\ref{q8c}), we obtain
the following integral representation for the potential:%
\begin{equation}
\Phi_{p,\nu}(\rho,z)=\widetilde{V}_{o}\int\limits_{0}^{1}\frac{\left(
1-u^{\prime2}\right)  ^{\nu}u^{\prime p+1}}{\sqrt{\left(  u+u^{\prime}\right)
^{2}+\overline{z}^{2}}}K\left[  \frac{4u^{\prime}u}{\left(  u+u^{\prime
}\right)  ^{2}+\overline{z}^{2}}\right]  du^{\prime} \ , 
\label{q8f}
\end{equation}
where the normalization constant, $\widetilde{V}_{o}$ is defined as:
\begin{equation}
\widetilde{V}_{o}=\left[  \frac{2(\nu+p/2+1)!}{\pi\left(  \frac{p}{2}\right)
!\nu!}\right]  V_{o} \ . 
\label{q8g}%
\end{equation}
The final expression in Eq. (\ref{q8f}) is general and provides a versatile
tool for analyzing the electrostatic potential of a wide class of axially
symmetric charge densities on a disk. This expression is valid for any real
values of $\nu>-1$ and $p>-1$ to ensure convergence of the integral. From both
analytical and computational perspectives, this formula is highly
advantageous. In particular, it allows for efficient numerical evaluation in
programming environments such as Python since all the functions involved are
readily available in scientific libraries. Moreover, the flexibility of the
parameters $\nu$ and $p$ makes this formulation highly adaptable to a broad
class of physically relevant charge distributions. It provides a unified
framework capable of capturing a range of profiles, from uniform to highly
peaked edge-concentrated densities.

\section{Discussions}

Some particular limits of Eq. (\ref{q8f}), which have already been discussed
in the literature, can be explicitly derived. The first limiting case
corresponds to a uniformly charged disk, characterized by the parameters
$p=\nu=0$. In this case, the electrostatic potential takes the form:%
\begin{equation}
\Phi_{p=0,\nu=0}(\rho,z)=\frac{2V_{o}}{\pi}%
\int\limits_{0}^{1}\frac{u^{\prime}}{\sqrt{\left(  u+u^{\prime}\right)
^{2}+\overline{z}^{2}}}K\left[  \frac{4u^{\prime}u}{\left(  u+u^{\prime
}\right)  ^{2}+\overline{z}^{2}}\right]  du^{\prime}, \label{q8i}%
\end{equation}
which has been discussed in Ref.\cite{el4}.

The second limit corresponds to a non-uniform surface charge distribution for
which the surface of the disk is an equipotential. This occurs when $p=0$ and
$\nu=-1/2$, yielding the potential:%
\begin{equation}
\Phi_{p=0,\nu=-1/2}(\rho,z)=\frac{V_{o}}{\pi}%
\int\limits_{0}^{1}\frac{\left(  1-u^{\prime2}\right)  ^{-1/2}u^{\prime}%
}{\sqrt{\left(  u+u^{\prime}\right)  ^{2}+\overline{z}^{2}}}K\left[
\frac{4u^{\prime}u}{\left(  u+u^{\prime}\right)  ^{2}+\overline{z}^{2}%
}\right]  du^{\prime}, \label{q8j}%
\end{equation}
which has been analyzed in Ref.\cite{el7}.
\begin{figure}[ptbh]
\begin{center}
\includegraphics[width=12cm]{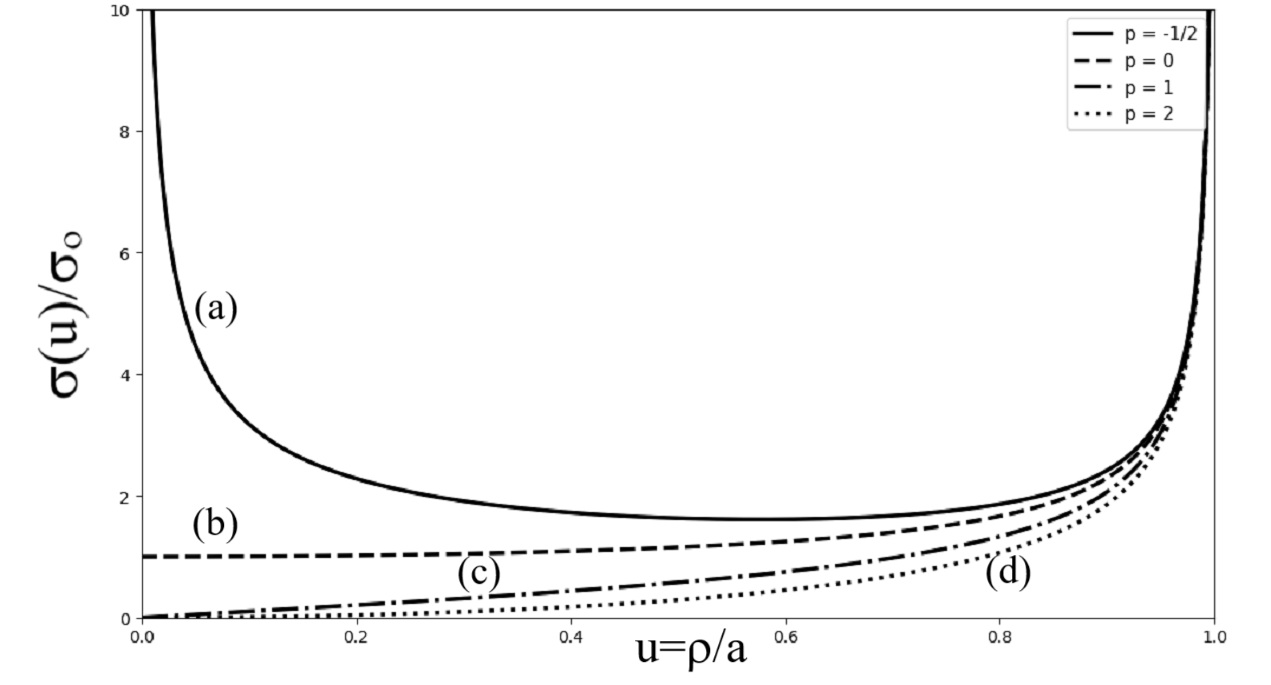}
\end{center}
\caption{ Behavior of the dimensionless surface charge density, $\sigma
(u)/\sigma_{o}$ as a function of the dimensionless radial distance, $u=\rho
/a$, for a non-uniform charge distribution given by $\sigma(u)=\sigma
_{o}\left(  1-u^{2}\right)  ^{\nu}u^{p}$. The parameter $\nu$ is fixed at
$\nu=-1/2$, while the values of the other parameter are chosen as
$p=-1/2,0,1$, and $2$. }%
\label{figone}%
\end{figure}

To illustrate the behavior of the non-uniform surface charge density defined
by Eq. (\ref{q4}), we plot in Figure 1 the dimensionless charge density
profile, $\sigma(u)/\sigma_{o}$, as a function of the dimensionless radial
coordinate $u=\rho/a$. In this figure, the parameter $\nu$ is fixed at
$\nu=-1/2$, while the parameter $p$ is varied over the values $p=-1/2,0,1$ and
$2$. For values $p\geq0$, the charge density exhibits a singularity at the
edge of the disk, i. e., at $u=1$. This indicates that the surface charge
accumulates strongly near the perimeter of the disk. In contrast, for values
$-1<p<0$, the distribution becomes singular both at the center ($u=0$) and the
edge ($u=1$) of the disk, reflecting a more complex concentration of charge
toward these regions. 

A particularly interesting case occurs when $\nu=-1/2$ and $p=0$.
In this limit, the resulting charge distribution corresponds to a
configuration in which the disk behaves as an \textit{equipotential
conductor}. The electrostatic potential on the surface of the disk (i. e., at
$z=0$, with $0\leq\rho\leq a$) is then constant and equal to $\pi V_{o}/4$
where $V_{o}=Q/(2\pi\varepsilon_{o}\,a)$. This equipotential case is often
used as a benchmark for validating analytical and numerical solutions in electrostatics.

The general expression in Eq. (\ref{q8f}) recovers well-known results for
specific surface charge distributions as special cases. For instance: i) When
$\nu=0$ and $p=0$, we recover the uniform charge distribution corresponding to
a constant surface density $\sigma_{o}=Q/(\pi a^{2})$; ii) When $\nu=-1/2$ and
$p=0$, we obtain a non-uniform distribution concentrated near the edge of the
disk resulting in a normalization constant, $\sigma_{o}=Q/(2\pi a^{2})$. These
limiting cases highlight how the general formulation not only broadens the
analytical scope of the problem, but also provides a powerful tool for
modeling realistic physical systems involving disk-like charge configurations.

To gain a deeper understanding of how different surface charge distributions
influence the resulting electrostatic potential, we undertake a detailed
analysis of Eq.(\ref{q8f}) by exploring various combinations of the parameters
$\nu$ and $p$, which characterize the functional form of the surface charge
density. In particular, we focus on the dimensionless potential on the plane
$z=0$ when $\nu=-1/2$. Such a potential is defined as $V_{p}(u)\equiv
\Phi_{p,\nu=-1/2}(\rho,0)/\Phi_{p,\nu=-1/2}(0,0)$ where $u=\rho/a$ is the
dimensionless radial distance. This particular choice of $\nu=-1/2$
corresponds to a family of non-uniform charge distributions that exhibit a
pronounced enhancement near the edge of the disk, a feature often observed in
practical electrostatic configurations involving finite conducting surfaces.

\begin{figure}[ptbh]
\begin{center}
\includegraphics[width=14cm]{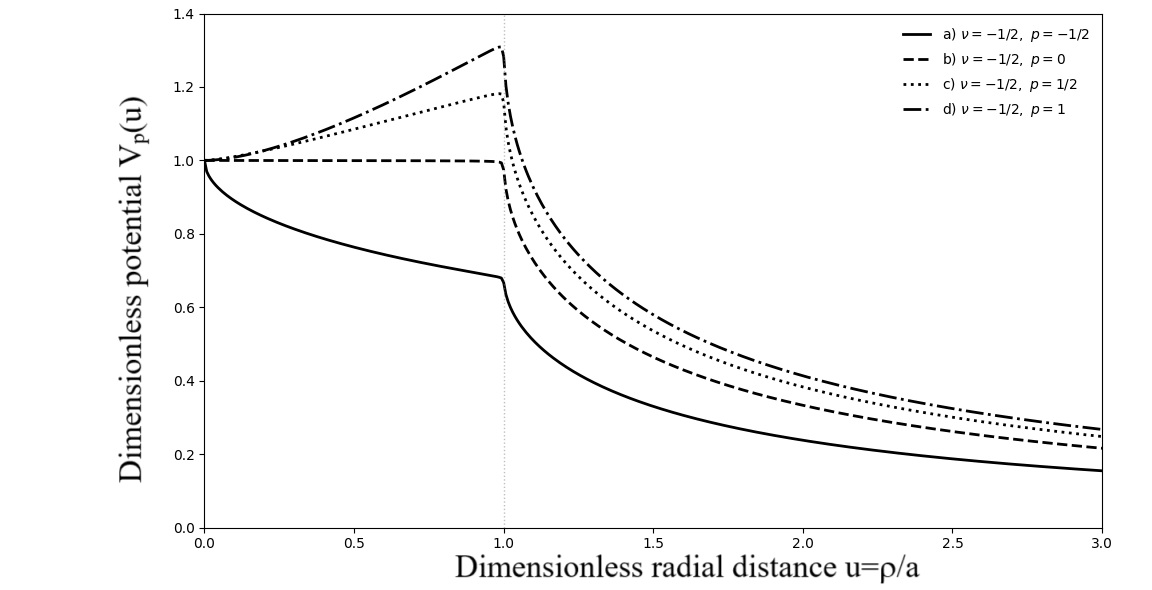}
\end{center}
\caption{ Dimensionless electrostatic potential in the plane $z=0$ of a
charged disk, $V_{p}(u)\equiv\Phi_{p,\nu=-1/2}(\rho,0)/\Phi_{p,\nu=-1/2}(0,0)$
as a function of dimensionless distance $u=\rho/a$. The disk has a non-uniform
surface charge density $\sigma(u)=\sigma_{o}\left(  1-u^{2} \right)  ^{\nu
}u^{p}$ where $\sigma_{o}$ is given by Eq. (\ref{q7}), and we consider the
following four cases: (a) $\nu=-1/2,p=-1/2$; (b) $\nu=-1/2,p=0$; (c)
$\nu=-1/2,p=1/2$; and (d) $\nu=-1/2,p=1$. }%
\label{figtwo}%
\end{figure}

In Figure 2, we present the behavior of $V_{p}(u)$ as a function of $u$ for
several representative values of the parameter $p$, which modulates the radial
profile of the surface charge density. Specifically, we consider the following
four cases: a) $\nu=-1/2,p=-1/2$; b) $\nu=-1/2,p=0$; c) $\nu=-1/2,p=1/2$; and
d) $\nu=-1/2,p=1$. Each case represents a distinct functional form of the
surface charge density allowing us to assess how changes in the radial
exponent $p$ affect the structure and spatial decay of the resulting
electrostatic potential $V_{p}(u)$.

By preserving the edge singularity induced by the value $\nu=-1/2$ and
adjusting the radial weighting through $p$, we obtain valuable insights into
the sensitivity of the potential field to variations in the charge
distribution, which is essential for modeling and interpreting electrostatic
interactions in systems with non-uniform axially-symmetric surface charge
distributions. This investigation offers valuable insights into the
sensitivity of the electrostatic potential to variations in the surface charge
distribution. This is of fundamental importance in the modeling of
electrostatic interactions in systems with radially inhomogeneous or
engineered charge profiles. The results not only highlight the role of the
radial exponent $p$, but also illustrate how different functional forms of
charge modulation can be used to tailor electrostatic potentials for specific applications.

From the analysis of Figure 2, we observe distinct behaviors of the
electrostatic potential $V_{p}(u)$ depending on the parameter $p$, which
characterizes the radial charge distribution on the disk. In the particular
case where $p=0$, the potential remains constant across the surface of the
disk, i. e., for $0\leq u\leq1$. This reflects the \textit{equipotential}
nature typical of a non-uniformly charged disk. Outside the disk, for $u>1$,
the potential decreases monotonically as the radial distance increases. When
$-1<p<0$, the potential behaves differently: $V_{p}(u)$ decreases
monotonically starting from the center of the disk ($u=0$) and continues to
decrease as $u$ increases. This indicates a higher concentration of charge
near the center, resulting in a maximum potential at the origin and a gradual
decline outward. In contrast, the potential exhibits a non-monotonic profile
for positive values of $p$. Initially, $V_{p}(u)$ increases with $u$, reaching
a maximum near the edge of the disk ($u\simeq1$), and then decreases
monotonically for $u>1$. This behavior is associated with charge distributions
that are more concentrated near the disk's boundary, leading to an enhanced
electric potential in the peripheral regions.

Another very interesting detail that we observed is that 
there exists a critical value of parameter, $p$ which we denoted as $p_c$
at which the behavior of $V_{p}(u)$ curve changes.
We estimated that $p_{c} \approx -0.34$.
More specifically speaking, for $p>p_{c}$, a maximum point is observed, 
while for $p<p_{c}$, the maximum is no longer present.

\section{Conclusions}

The potential of a charged disk with a radially symmetric surface charge
density is a critical concept in electrostatics, as it provides valuable
insight into the behavior of electric fields in systems with radial symmetry.
By determining the potential, one can calculate the resulting electric field
at any point in space, which is essential for understanding forces acting on
other charges within the field. The importance of finding this potential lies
in its ability to simplify complicated problems where the charge distribution
exhibits symmetry, such as in capacitors or disk-shaped electrodes. Knowing
the potential allows engineers and physicists to design and analyze
electrostatic systems with precision, optimize the performance of devices, and
predict interactions between charged objects. Moreover, the solution to the
potential can often be extended or adapted to more complicated geometries or
varying charge distributions, making it a foundational tool in electrostatic theory.


In this work, we have derived an exact expression for the electrostatic
potential generated by a charged disk with non-uniform surface charge density
that is radially symmetric. We consider a general surface charge distribution
of the form, $\sigma(u)=\sigma_{o}\left(  1-u^{2}\right)  ^{\nu}u^{p}$
described by two real parameters ($\nu>-1$ and $p>-1$) where $u=\rho/a$ is a
dimensionless 2D radial distance measured relative to the center of the disk.
The relevant expression for the electrostatic potential can be given in
compact form in terms of an integral of Bessel functions. By focusing our
attention on the behavior of the potential on the plane of the disk, we
analyzed its properties as a function of the radial exponent, $p$ while
keeping the edge singularity fixed at the value, $\nu=-1/2$. To this effect,
we looked at the properties of the in-plane potential in dimensionless form,
$V_{p}(u)\equiv\Phi_{p,\nu=-1/2}(\rho,0)/\Phi_{p,\nu=-1/2}(0,0)$. The results
reveal three characteristic regimes: (i) A flat, constant potential across the
disk for $p=0$; (ii) A monotonically decreasing potential from the center for
$-1<p<0$, reflecting center-concentrated charge; and (iii) A non-monotonic
profile of the potential for $p>0$, with the potential peaking near the disk
edge due to boundary-enhanced charge density. These findings illustrate how
tailoring the charge profile influences the spatial structure of the potential
and provide essential insights for applications involving engineered surface
charges in electrostatic systems.




\section*{Acknowledgements}

The research of O. Ciftja was supported in part by National Science Foundation
(NSF) Grant No. DMR-2001980.



\begin{thebibliography}{9}                                                                                                %

\bibitem {6}D. J. Griffiths, \textit{Introduction to Electrodynamics}
(Cambridge University Press, Cambridge, 2017).

\bibitem {7}J. R. Reitz and F. J. Milford, \textit{Foundations of
Electromagnetic Theory} (Addison-Wesley, Reading, 1960).

\bibitem {8}J. D. Jackson, \textit{Classical Electrodynamics} (John Wiley \&
Sons, Nova Jersey, 1999).

\bibitem {9}J. A. Stratton, \textit{Electromagnetic Theory} (John Wiley \&
Sons, Hoboken, 2007).

\bibitem {10}W. K. H. Panofsky and M. Phillips, \textit{Classical Electricity
and Magnetism} (Addison-Wesley, Reading, 1962).

\bibitem {11}L. D. Landau, E. M. Lifshitz and L. P. Pitaevskii,
\textit{Eletrodynamics of Continuous Media}, Vol. 8 of Course of Theoretical
Physics (Pergamon Press, Oxford, 1984).

\bibitem {12}L. Eyges, \textit{The Classical Electrodynamic Field} (Dover, New
York, 2012).

\bibitem {13}E. J. Konopinski, \textit{Eletromagnetic Fields and Relativistic
Particles} (McGraw-Hill, New York, 1981).

\bibitem {el1}F. R. Zypman, \textit{Off-axis Electric Field of a Ring of
Charge}, \emph{Am. J. Phys}. \textbf{74}, 295 (2006).

\bibitem {el2}S. Datta, \textit{Electric and Magnetic Fields from a Circular
Coil using Elliptic Integrals}, \emph{Phys. Educ. India} \textbf{24}, 203 (2007).

\bibitem {el3}O. Ciftja, A. Babineaux and N. Hafeez, \textit{The Electrostatic
Potential of a Uniformly Charged Ring}, \emph{Eur. J. Phys.} \textbf{30}, 623 (2009).

\bibitem {el4}O. Ciftja and I. Hysi, \textit{The Electrostatic Potential of a
Uniformly Charged Disk as the Source of Novel Mathematical Identities},
\emph{Appl. Math. Lett.} \textbf{24}, 1919 (2011).

\bibitem {el5}V. Bochko and Z. K. Silagadze, \textit{On the Electrostatic
Potential and Electric Field of a Uniformly Charged Disk}, \emph{Eur. J.
Phys.} \textbf{41}, 045201 (2020).

\bibitem {el6}F. Escalante, \textit{Electrostatic Potential and Electric Field
in the }$\mathit{z}$\textit{\ Axis of a non Centred Circular Charged Ring},
\emph{Eur. J. Phys.} \textbf{42}, 065703 (2021).

\bibitem {el7}O. Ciftja, \textit{Results for charged disks with differente
forms of surface charge density}, \emph{Results Phys}. \textbf{16}, 102962 (2020).

\bibitem {el8}A. E. Sagaydak and Z. K. Silagadze, \textit{Electrostatic
potential of a uniformly charged disk through Green's theorem}, \emph{Eur. J.
Phys. }\textbf{46}, 015203 (2025).

\bibitem {g1}T. Morgan and L. Morgan, \textit{The Gravitational Field of a
Disk}, \emph{Phys. Rev.} \textbf{188}, 2544 (1969); Errata: \emph{Phys. Rev.
D} \textbf{1}, 3522 (1970).

\bibitem {g2}F. T. Krogh, E. W. Ng and W. V. Snyser, \textit{The Gravitational
Field of a Disk}, \emph{Celestial Mech}. \textbf{26}, 395 (1982).

\bibitem {g3}H. Lass and L. Blitzer, \textit{The Gravitational Potential Due
to Uniform Disks and Rings}, Celestial Mech. \textbf{30}, 225 (1983).

\bibitem {g4}G. Neugebauer and R. Meinel, \textit{General Relativistic
Gravitational Field of a Rigdly Rotating Disk of Dust: Axis Potential, Disk
Metric, and Surface Massa Density}, \emph{Phys. Rev. Lett.} \textbf{73}, 2166 (1994).

\bibitem {g5}J. T. Conway, \textit{Analytical Solution for the Newtonian
Gravitational Field Induced by Matter within Axisymmetric Boudaries},
\emph{Mon. Not. R. Astron. Soc}. \textbf{316}, 540 (2000).

\bibitem {g6}A. Elipe, E. Tresaco and A. Riaguas, \textit{Gravitational
Potential of a Massive Disk}, \emph{Adv. Astr. Sci}. \textbf{143}, 843 (2009).

\bibitem {g7}J. M. Hur\'{e} e F. Hersant, \textit{The Newtonian Potential of
Thin Disks}, \emph{Astr. Astrophys.} \textbf{531}, A6 (2011).

\bibitem {g8}E. Schulz, \textit{The Gravitational Force and Potential of the
Finite Mestel Disk}, \emph{The Astrophys. J.} \textbf{747}, 106 (2012).

\bibitem {g9}J. D. Weiss, \textit{Certain Aspects of the Gravitational Field
of a Disk}, \emph{Appl. Math.} \textbf{9}, 1360 (2018).

\bibitem {g10}A. J. Silenko and Y. A. Tsalkou, \textit{Quasi-Uniform
Gravitational Field of a Disk Revisited}, \emph{Int. J. Mod. Phys. A}
\textbf{34}, 1950228 (2019).

\bibitem {n1}B. Tiang, X. Zheng, T. J. Kempa, Y. Fang, N. Yu, G. Yu, J. Huang,
and C. M. Lieber, \textit{Coaxial Nanowires as Solar Cells and Nanoelectronic
Power Sources}, \emph{Nature} \textbf{449}, 885 (2007).

\bibitem {n2}A. Nduwimana and X. -Q. Wange, \textit{Charge Carrier Separation
in Modulation Doped Coaxial Semiconductor Nanowires}, \emph{Nano. Lett.
}\textbf{9}, 283 (2009).

\bibitem {b1}Y. Levin, \textit{Electrostatic Correlations: from Plasma to
Biology}, \emph{Rep. Prog. Phys}. \textbf{65}, 1577 (2002).

\bibitem {b2}A. L. Kholodenko and A. L. Beyerlein,\textit{\ Time-Dependent
Density-Functional Theory for Multicomponent Systems}, \emph{Phys. Rev. A}
\textbf{43}, 3309 (1986).

\bibitem {b3}R. D. Coalson and A. Duncan, \textit{Systematic Ionic Screening
Theory of Macroions}, \emph{J. Chem. Phys.} \textbf{97}, 5653 (1992).

\bibitem {b4}A. Agra, E. Trizak and L. Bocquet, \textit{The Interplay Between
Screening Properties and Colloid Anisotropy: Towards a Reliable pair Potential
for Disc-Like Charged Particles}, \emph{Eur. Phys. E} \textbf{15}, 345 (2004).

\bibitem {f1}R. Morf and B. I. Halperin, \textit{Monte Carlo Evaluation of
Trial Wave Functions for the Fractional Quantized Hall Effect: Disk Geometry},
\emph{Phys. Rev. B} \textbf{33}, 2221 (1986).

\bibitem {f2}J. Xia,\textit{\ An Estimate of the Ground State Energy of the
Fractional Quantum Hall Effect}, \emph{J. Math. Phys}. \textbf{40}, 150 (1999).

\bibitem {f3}O. Ciftja, \textit{Monte Carlo Study of Bose Laughlin Wave
Function for Filling Factors }$1/2$\textit{, }$1/4$\textit{\ and }$1/6$,
\emph{Europhys. Lett}. \textbf{74}, 486 (2006).

\bibitem {f4}O. Ciftja, \textit{Exact Results for Systems of Electrons in the
Fractional Quantum Hall Regime II} , \emph{Physica }B \textbf{404} 2244
(2009); Errata: \emph{Physica B} \textbf{406}, 2054 (2011).

\bibitem {f5}O. Ciftja, \textit{Analytic Wave Functions for the Half-Filled
Lowest Landau Level}, \emph{Int. J. Mod. B} \textbf{24}, 3489 (2010).

\bibitem {18}G. B. Arfken and H. J. Weber, \textit{Mathematical Methods for
Physicists}, 3$^{a}$ edi\c{c}\~{a}o (Academic Press, California, 1985).

\bibitem {be1}N. J. Sonine, \textit{Recherches sur les fonctions cylindriques
ei le d\'{e}veloppement des fonctions continues en s\'{e}ries}, \emph{Math.
Ann}. \textbf{16}, 1 (1880).

\bibitem {be2}I. S. Gradshteyn and I. M. Ryzhik, \textit{Table of Integrals,
Series and Products}, Academic Press, New York, 2007.

\bibitem {be3}P. F. Byrd and M. D. Friedman, \textit{Handbook of Elliptic
Integrals for Engineers and Scientists}, 2nd ed., Springer-Verlag, New
York--Berlin, 1971.
\end{thebibliography}
\end{document}